\documentclass[a4paper]{jpconf}
\usepackage{graphicx}
\begin{document}
\title{Targeted searches for gravitational waves from radio pulsars}

\author{R\'ejean J. Dupuis for the LIGO Scientific Collaboration}

\address{California Institute of Technology, Pasadena, CA 91125, USA}

\ead{rejean@caltech.edu}

\begin{abstract}
An overview of the searches for gravitational waves from radio pulsars with LIGO and GEO is given.   We give a brief description of the algorithm used in these targeted searches and provide end-to-end validation of the technique through hardware injections.  We report on some aspects of the recent S3/S4 LIGO and GEO search for signals from several pulsars. The gaussianity of narrow frequency bands of S3/S4 LIGO data, where pulsar signals are expected,  is assessed with Kolmogorov-Smirnov tests.  Preliminary results from the S3 run with a network of four detectors are given for pulsar J1939+2134.
\end{abstract}

\section{Introduction}
Radio pulsars are interesting sources for currently operating gravitational wave detectors~\cite{cc,bo}.  The positions and spin evolutions of these sources are well known from radio observations which reduce the parameter space considerably compared to all-sky searches for unknown neutron stars.  In addition to simplifying the data analysis problem, the smaller parameter space in these \emph{targeted searches} improves the sensitivity to  gravitational waves by lowering the detection threshold.

LIGO and GEO have so far conducted two searches for gravitational waves from radio pulsars using data from the S1 science run~\cite{ab1} and data from the S2 run~\cite{ab2}.  Neither of these searches provided a detection but upper limits were set on gravitational wave emission for selected pulsars. Both of these analyses searched for gravitational waves  at twice the rotational frequency of pulsars, where we would expect a signal from an asymmetric pulsar. 

In the S1 analysis, two techniques were used to set upper limits on gravitational wave emission from pulsar J1939+2134 (the fastest rotating known millisecond pulsar)~\cite{ab1}: a Bayesian time-domain method and a classical frequency-domain method.   The main result from this S1 analysis was an upper limit on signals from pulsar J1939+2134 of $h_0 < 1.4\times10^{-22}$ with 95\% confidence.

 Here we will focus on the time-domain Bayesian method which is more suited for targeted sources~\cite{dw}.  The frequency-domain statistical technique used in the S1 paper is currently being used for broad all-sky searches for unknown sources, including Einstein@home~\cite{jks,EaH}.

For the LIGO S2 run, the time-domain search was expanded to include all well-known isolated pulsars with putative gravitational wave frequencies above 40\,Hz~\cite{ab2}.  Using the S2 data, multi-detector upper limits were set on gravitational wave emission from 28 pulsars including J1939+2134  and the Crab pulsar. The tightest limit on gravitational wave strain came from pulsar
J1910-5959D with a 95\% upper limit of $h_0 < 1.7\times 10^{-24}$.  At the
time of S2 this was the lowest upper limit ever set on a pulsar by a gravitational wave
detector.   The four best limits on equatorial ellipticity for S2 came from
the four closest pulsars J0030+0451, J1024$-$0719, J1744$-$1134, and
J2124$-$3358 with 95\% upper limits on the ellipticity of
$4.8\times 10^{-6}$, $8.6\times 10^{-6}$, $8.3\times 10^{-6}$, and $4.5\times 10^{-6}$,
respectively.

The analysis of S3/S4 LIGO and GEO data is currently underway with more sensitive data and  a larger selection of pulsars than in S2. The main change in our search since S2 has been the addition of pulsars in binary systems.  The timing for all sources was supplied by the Pulsar Group at Jodrell Bank Observatory. 

\begin{figure}
\begin{center}
\includegraphics[width=8cm]{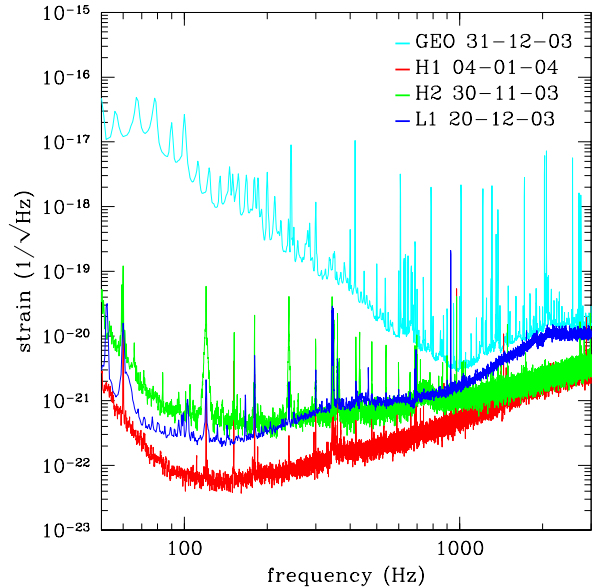}
\end{center}
\caption{\label{S3sensitivity}Best sensitivities of the three
LIGO and the GEO\,600 interferometers during the S3 data
run.}
\end{figure}

We are currently analyzing S3/S4 data for gravitational waves from all known radio pulsars with $f_{\rm GW}>40~{\rm Hz}$ for which the source parameters are known sufficiently well.  This breaks down to a  total of 93 pulsars of which 60 are binaries and 33 are isolated.  The improved sensitivity of the detectors in S3/S4 promises to give interesting results for several sources.  For the Crab pulsar, we should be within a factor of a few of the  spin-down based upper limit~\cite{ab1}.  Figure~\ref{S3sensitivity} shows the sensitivity of the GEO and LIGO interferometers during the S3 run.

In this paper we will report on some aspects of the new S3/S4 search for signals from radio pulsars.  In Section~\ref{method} we will give a brief review of the technique used in previous and current analyses.  We will provide an example of a hardware injection of a pulsar signal in the LIGO interferometers during S3 in Section~\ref{injections}.  In Section~\ref{S3S4data} we will show the results of statistical tests investigating the gaussianity of the frequency bands of the LIGO S3/S4 data containing the radio pulsar signals.    As an example, we will show preliminary results in Section~\ref{prelim} for a multi-detector analysis for pulsar J1939+2134 using LIGO and GEO S3 data.  Brief conclusions will be given in Section~\ref{conclusions}.

\section{Bayesian time-domain method}
\label{method}
The overall pipeline for the analysis of the S3/S4 data for signals from known pulsars is very similar to the techniques used in S1 and S2.  A brief description of the method is provided here with more details available in~\cite{dw,rjd}. 

The data from the interferometer are heterodyned with the known phase evolution of each pulsar and down-sampled to one data point, $B_k$, each $k$th minute. These data points $B_k$ are essentially a 1/60\,Hz band-limited time series centered on the instantaneous frequency of the gravitational wave signal at the detector.

We assume that the noise floor for each band was stationary and gaussian over periods of 30 minutes. As will be shown later in Section~\ref{S3S4data}, this is a reasonable assumption for a
majority of the pulsars.  

We take a Bayesian approach for the statistical analysis.  We use
Bayes' Theorem to calculate the posterior probability,
$p(\textrm{\bf a}|\{B_k\})$, of a set of parameters ${\bf a}$ given
the binned data, $\{B_k\}$. For this problem, Bayes' Theorem states
that
\begin{equation}
p(\textrm{\bf a}|\{B_k\}) = \frac{p(\textrm{\bf
a})p(\{B_k\}|\textrm{\bf a})}{p(\{B_k\})},
\end{equation}
where $\textrm{\bf a}$ represents the set of parameters that could
produce the set of data, $\{B_k\}$, with likelihood
$p(\{B_k\}|\textrm{\bf a})$.   Note that $\textrm{\bf a}$ comprises of four unknown parameters:
the gravitational 
wave amplitude  $h_0$, the polarization
angle  $\psi$, the inclination of the
pulsar's spin axis with respect to the line-of-sight $\iota$, and  the initial phase
of the gravitational wave signal $\phi_0$.  Our prior beliefs in our set of
parameters are reflected in the prior probability term,
$p(\textrm{\bf a})$. We use the least informative priors for
most of the parameters in their respective ranges: $\phi_{0}$
uniform over $[0,2\pi]$, $\psi$ uniform over $[-\pi/4,\pi/4]$, and
$\iota$ uniform in $\cos\iota$ over $[-1,1]$, corresponding to a
uniform prior per unit solid angle of pulsar orientation.

It can be shown that the likelihood of each 30 minute segment of data, labeled by $j$, is given by
\begin{equation}
\label{likelihood}
p(\{B_k\}_j|{\mathbf a}) \propto
 \left(\sum_{k=k_{1(j)}}^{k_{2(j)}}|B_k-y_k|^2\right)^{-m_j},
\end{equation}
where $y_k$ is the signal model, $B_k$ is the are the data points, and $m_j = 30$ is the number of 
data points in the segment.  
Equation~\ref{likelihood} is equivalent to a Student's $t$-distribution with $2m_j -1$
degrees of freedom.  We note that the lengths of the stationary segments, $m_j$, can be adjusted
depending on the performance of the detectors.

The joint likelihood of all the $M$ stretches of data, taken as
independent, is therefore
\begin{equation}
p(\{B_k\}|\mathbf{a}) \propto  \prod_j^M p(\{B_k\}_j|\mathbf{a}).
\label{StudentTL}
\end{equation}

\section{Hardware injections}
\label{injections}

We use the term \emph{hardware injections} for
simulated signals that make the instrument behave in the same way as if a
gravitational wave signal was present.  During the S3 and S4 runs, several
hardware injections were carried out that mimicked periodic signals
from pulsars in the LIGO interferometers.  These injections provide end-to-end validations of the search codes and the data
acquisition pipelines. The phase of the signals is properly modulated to simulate the
Doppler shift from specific directions in the sky.  The signals are also amplitude modulated to 
reflect the antenna pattern of the interferometers.

These pulsar hardware injections increase our confidence
that the timing between sites is consistent. This is especially
important for coherent multi-detector searches where a large lag in
timing between the detectors could severely reduce the sensitivity
of the search. 

\begin{figure}
\begin{center}
\includegraphics[width=8cm]{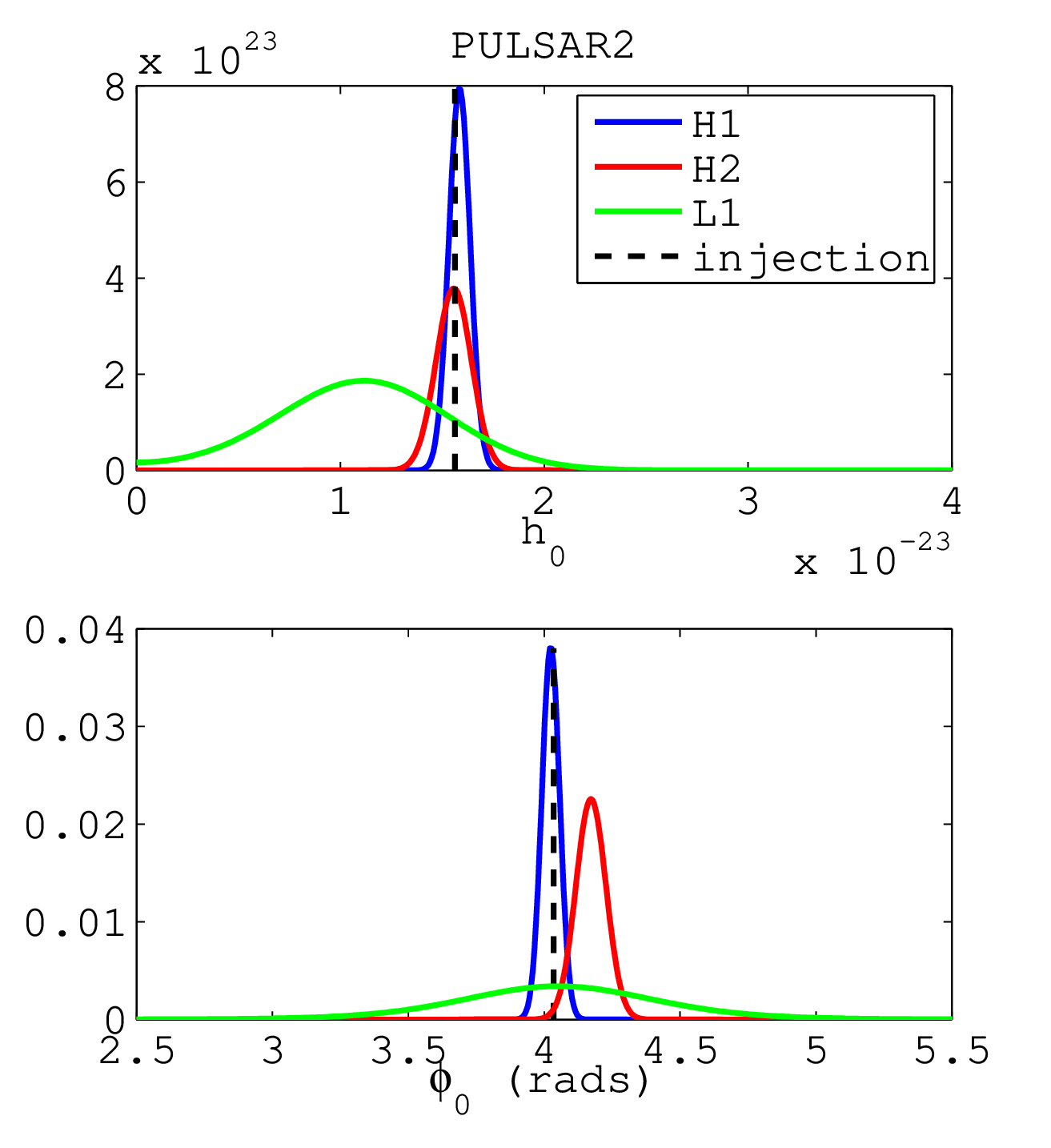}
\end{center}
\caption{\label{S3PULSAR2} Recovered parameters for one of the S3
pulsar hardware injections. The dashed line represents the
signal parameter injected. }
\end{figure}

Figure~\ref{S3PULSAR2} shows the marginalised probability distribution functions for $h_0$ and $\phi_0$ for one of the S3 pulsar injections.  The recovered parameters for this example agree well with the injected signal parameters.  Work is currently in progress to make sure all the S3/S4 hardware injections are correctly extracted from the LIGO data.  There are still some unclear issues with the phase of the S4 injections but we do not expect any major problems.

\section{Characterizing the S3/S4 LIGO data}
\label{S3S4data}

 In this section we report on some statistical tests that were performed to assess the gaussianity of the S3/S4 data near the frequencies where we expect pulsar signals.  For each pulsar, we examined the data in a $1/60$\,Hz frequency band around the instantaneous gravitational wave frequency of the signal at the detector.  These data are the heterodyned data set mentioned earlier.  

We have applied Kolmogorov-Smirnov (Lilliefors) tests to assess whether contiguous segments of 30 minutes of this data are normally distributed for each pulsar.  This consists of comparing the experimental cumulative distribution function (cdf) for each of the 30 minute data segments with a gaussian cdf. The K-S test statistic is a simple method to quantitatively measure the difference of these two distributions.  

\begin{figure}
\begin{center}
\includegraphics[width=7cm]{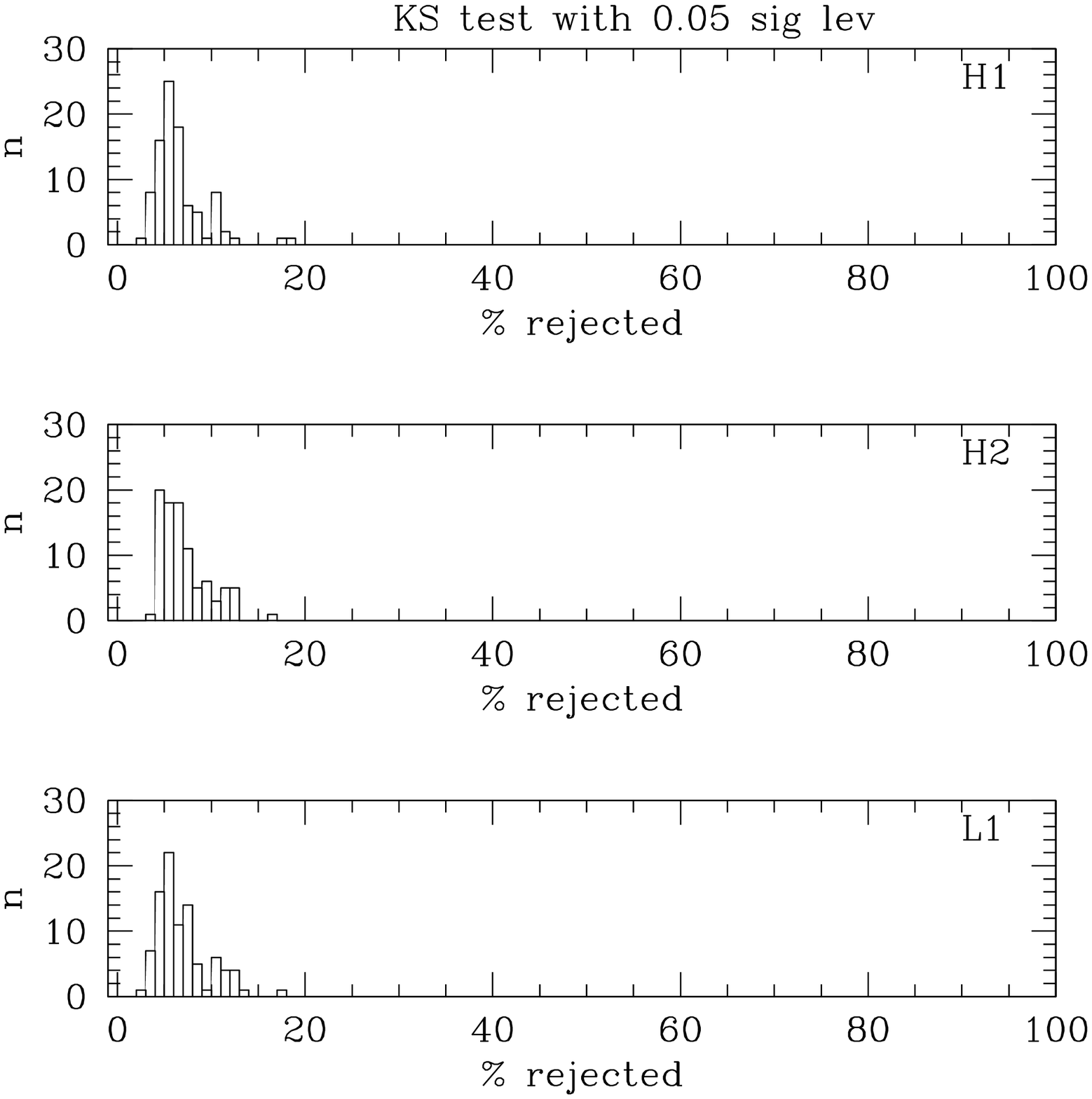} \ \ \  \  \ \
\includegraphics[width=7cm]{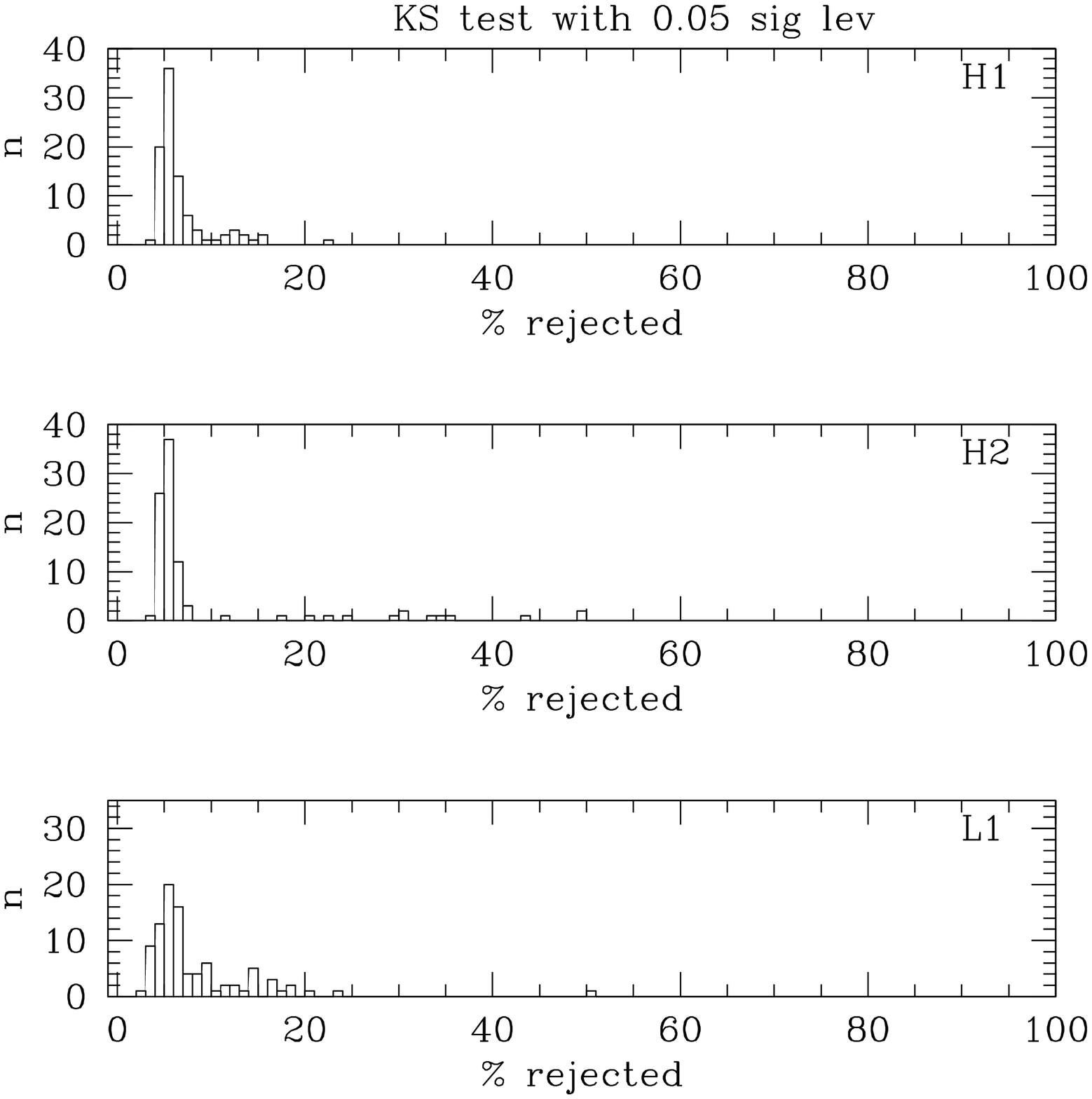}
\end{center}
\caption{\label{KS}  Histogram of the percentage of data which is rejected at a 0.05 significance
level with a Kolmogorov-Smirnov (Lilliefors) test for
normality for each of the 93 pulsars with the S4 (left) and S3 (right) data.}
\end{figure}

For each pulsar, the K-S statistic was calculated for every 30 minute
segment of data and compared to the Lilliefors critical values~\cite{Lilliefors}.
The critical value for a sample size of 60 (since the $B_k$'s are complex) for a level of significance of 0.05 is $D = 0.114$.  By definition, at a level of significance of 0.05 we would expect
approximately 5\% of the data to be rejected if the data was indeed
normal. Figure~\ref{KS} shows the distribution of the fraction
of 30 minute segments of data rejected for each interferometer.  For
the majority of the pulsars between 5-10\% of the data were
rejected. This is quite reasonable and suggests that the data is
relatively well behaved for most pulsars.   For a few pulsars the S4 data
is significantly better modeled by a gaussian distribution than the S3 data.  This is due to an 
improvement in the performance of the detectors in S4.  It has been shown that rejecting the 
non-gaussian S3 data does not significantly effect the final results~\cite{rjd}.  This is not surprising since
the non-gaussian data will have a small likelihood.

\section{Preliminary multi-detector analysis between LIGO and GEO}
\label{prelim}
\begin{figure}
\begin{center}
\includegraphics[width=8cm]{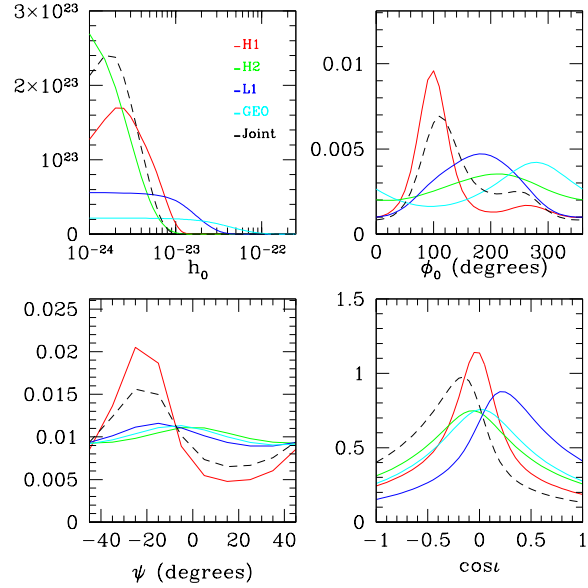}
\end{center}
\caption{\label{193721} Preliminary posterior probability distributions for
pulsar J1939+2134 using S3 data from GEO and the three LIGO interferometers.}
\end{figure}

The full and final results from the S3/S4 LIGO and GEO targeted pulsar search will be presented in a future paper by the LIGO Scientific Collaboration and Jodrell Bank Observatory.  Here we present preliminary results for pulsar J1939+2134 using GEO and LIGO S3 data.  Figure~\ref{193721} shows the posterior probability distributions for parameters of a signal from J1939+2134 using the S3 data.  It is clear that no gravitational wave signals are detected.   It can be seen from the plot that an 95\% upper limit on gravitational wave strain from J1939+2134 using S3 data can be set to $\sim 5 \times 10^{-24}$.  This limit is about a factor of two better than the S2 results~\cite{ab2} and a factor of 20 better than the S1 result~\cite{ab1}.  We can expect further improvement when the S4 data is added to the analysis in the near future.

\section{Conclusions}
\label{conclusions}
Upper limits on gravitational wave emission from a selection of 28 isolated pulsars were set using the LIGO S2 data~\cite{ab2}.  A similar analysis, including binary pulsars, is currently underway using the S3/S4 LIGO and GEO data.  This analysis will include 93 radio pulsars and is nearly complete. 

In this paper we have investigated the gaussianity of the LIGO S3/S4 data in narrow frequency bands where the pulsar signals are expected.  Generally the data are well described by a gaussian distribution as expected. We have presented a preliminary multi-detector upper limit for pulsar J1939+2134 using LIGO and GEO S3 data which should be significantly  improved when the S4 data is included.

\ack
The authors gratefully acknowledge the support of the United States National 
Science Foundation for the construction and operation of the LIGO Laboratory 
and the Particle Physics and Astronomy Research Council of the United Kingdom, 
the Max-Planck-Society and the State of Niedersachsen/Germany for support of 
the construction and operation of the GEO600 detector. The authors also 
gratefully acknowledge the support of the research by these agencies and by the 
Australian Research Council, the Natural Sciences and Engineering Research 
Council of Canada, the Council of Scientific and Industrial Research of India, 
the Department of Science and Technology of India, the Spanish Ministerio de 
Educacion y Ciencia, the John Simon Guggenheim Foundation, the Leverhulme Trust,
 the David and 
Lucile Packard Foundation, the Research Corporation, and the Alfred P. Sloan 
Foundation.  This document has been assigned LIGO Laboratory document 
number LIGO-P050037-00-E.

\end{document}